Handreichung für Lehrpersonen:

# Die Bahnen der ISS und anderer Satelliten

### Klassen 9 – 13

**Markus Nielbock**

22. Januar 2019

## Zusammenfassung

Künstliche Satelliten eignen sich als anwendungsnahe Beispiele, um die Wirkung der Gravitation auf ihre Bahn näher zu erläutern. Dieses Arbeitsmaterial behandelt dazu exemplarisch den Orbit der Internationalen Raumstation (ISS) um die Erde. In einfachen Rechnungen und Darstellungen erfahren die Schülerinnen und Schüler, wie Bahnen künstlicher Satelliten zustande kommen und welche charakteristischen Geschwindigkeiten dabei auftreten. Sie vergleichen ihre Ergebnisse zur ISS mit denen von geostationären Satelliten und entdecken Anwendungen dieses besonderen Orbits.

## Lernziele

Die Schülerinnen und Schüler

- berechnen Satellitenorbits anhand von vereinfachten Annahmen,

- wenden Gleichungen der Mechanik mit mehreren Variablen an,

- können erläutern, welche Kräfte auf die ISS und die Astronauten wirken, die stabile Orbits erzeugen.

## Materialien

- Arbeitsblätter (erhältlich unter: http://www.haus-der-astronomie.de/raum-fuer-bildung)
- Stift
- Taschenrechner
- Lineal
- Papier
- Smartphone oder Computer mit Internetverbindung für Videos und Recherchen (optional)

## Stichworte

Raumstation, ISS, Orbit, Satelliten, Gravitation, Zentrifugalkraft, Zentripetalkraft

## Dauer

180 Minuten

# Hintergrund

## Künstliche Satelliten

Seit dem Start von Sputnik am 4. Oktober 1957 werden regelmäßig künstliche Satelliten und Sonden ins Weltall gebracht. Ihre Funktionen sind mit der Zeit immer komplexer geworden. Die Daten der Wetter- und Navigationssatelliten sind heute allgegenwärtig. Zudem helfen Erdbeobachtungssatelliten bei der Bewältigung wichtiger Aufgaben wie Katastrophenmanagement und Klimaüberwachung.

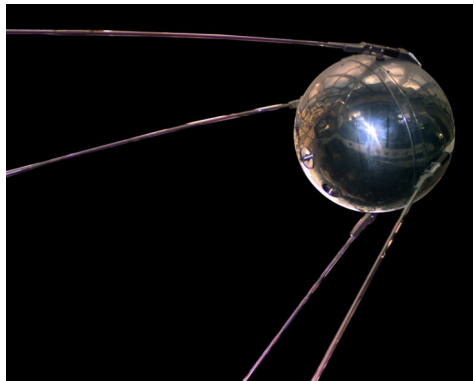

**Abbildung 1:** Nachbau des Sputnik 1 (NASA, https://nssdc.gsfc.nasa.gov/planetary/image/sputnik_asm.jpg).

## Die Internationale Raumstation

Die Internationale Raumstation (Abb. 2) ist ein internationales Projekt mit derzeit 15 beteiligten Nationen (ESA 2013; Garcia 2018a). Sie dient als wissenschaftliches Forschungslabor für Fragestellungen, deren experimentelle Untersuchung durch den Einfluss der Gravitation auf der Erde erschwert wird.

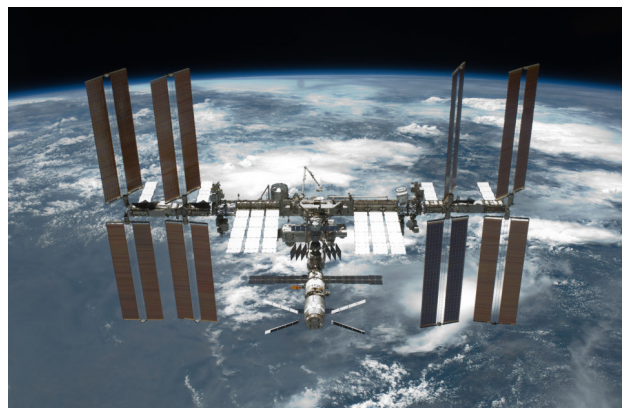

**Abbildung 2:** Die ISS im Jahre 2011 (Bild: NASA).

Neben der Materialforschung und biologischen Studien spielt auch die Medizin eine wichtige Rolle. Der Einfluss der Mikrogravitation führt zu Symptomen, die Krankheitsbildern auf der Erde ähneln. Daher hofft man, Erkenntnisse in der kontrollierten Umgebung der Raumstation zu erlangen, die auch bei der Erforschung der Krankheiten helfen und Therapien den Weg ebnen. Ein weiterer Grund besteht darin, langfristige Missionen innerhalb des Sonnensystems vorzubereiten.

Seit 1998 wird die ISS aufgebaut (Loff 2015) und mittels einzelner Module (Abb. 3) ständig erweitert (Zak 2017). Ihr Betrieb ist bis mindestens 2024 vorgesehen, wahrscheinlich aber sogar bis 2028 möglich (Sputnik 2016; Ulmer 2015). Die gesamte Struktur hat eine Masse von 420 t. Sie ist 109 m lang, 73 m breit (Garcia 2018b) und 45 m hoch (ESA 2014). Auf einer Bahnhöhe von etwa 400 km benötigt die ISS für eine Erdumrundung ungefähr 92 Minuten (Howell 2018).

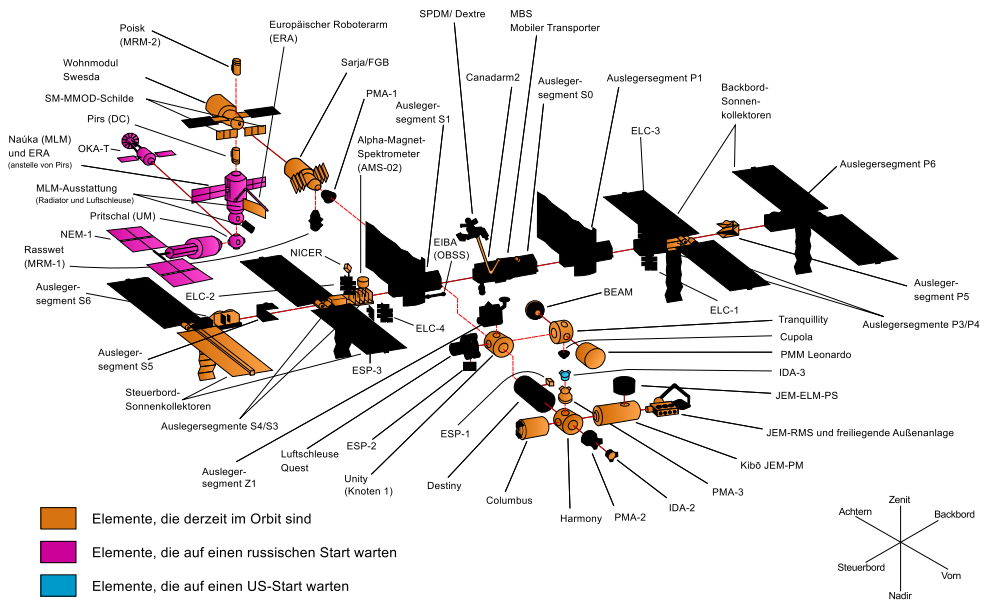

**Abbildung 3:** Die Module der ISS im Juni 2017 (Bild: NASA).

## Satellitenbahnen

Bahnen von künstlichen Erdsatelliten werden im wesentlichen durch drei Größen bestimmt. Das sind die Höhe über der Erdoberfläche, die Inklination und die Exzentrizität. Die Inklination ist der Winkel der die Bahn gegenüber dem Erdäquator einnimmt. Die Exzentrizität gibt an, wie sehr die Bahn von einem Kreis abweicht.

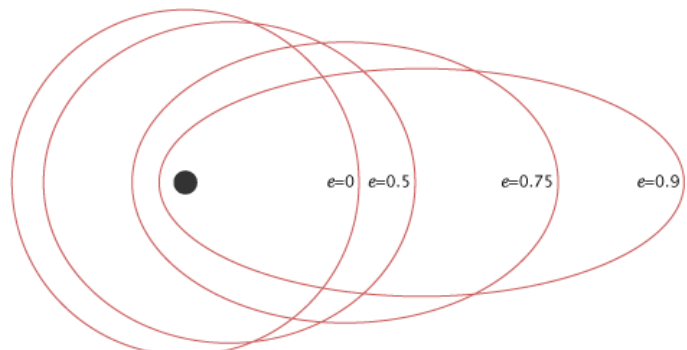

**Abbildung 4:** Ellipsen mit unterschiedlichen Exzentrizitäten (Bild: NASA/R. Simmon).

Alle Bahnen von Erdsatelliten sind sogenannte Ellipsen, wobei der Kreis ein Spezialfall einer Ellipse ist, deren Exzentrizität Null ist (Abb. 4). Die Bahn der ISS hat eine sehr geringe Exzentrizität, d. h.

sie ist nahezu perfekt kreisförmig. Sehr exzentrische Orbits verwendet man nur für spezielle Anwendungen. Bahnen mit großer Exzentrizität findet man jedoch oft bei Kometen. Das sind natürliche Objekte, die vom Rand des Sonnensystems in die Nähe der Sonne wandern.

Die Höhen der Orbits über dem Erdboden teilt man grob in vier Bereiche ein (Tab. 1).

**Tabelle 1:** Liste von Orbits nach Höhe über der Erde (Ichoku 2009).

| Abkürzung | Bezeichnung | Übersetzung | Höhe (km) |
|-----------|-------------|-------------|-----------|
| LEO | Low Earth Orbit | Niedriger Erdorbit | $180 - 2000$ |
| MEO | Mid Earth Orbit | Mittlerer Erdorbit | $2000 - 35780$ |
| GSO | Geosynchronous Orbit | Erdsynchroner Orbit | im Mittel 35786 |
| GEO | Geostationary Orbit | Geostationärer Orbit | 35786 |
| HEO | High Earth Orbit | Hoher Erdorbit | $> 35786$ |

Dazwischen gibt es noch einige feinere Unterteilungen, wie beispielsweise der Parkorbit, auf dem sich ein Raumschiff auf dem Weg zur ISS zunächst aufhält, bevor es langsam zur Raumstation aufsteigt. Die ISS befindet sich daher mit ihrer Bahn in einer Höhe von etwa 400 km in einem LEO.

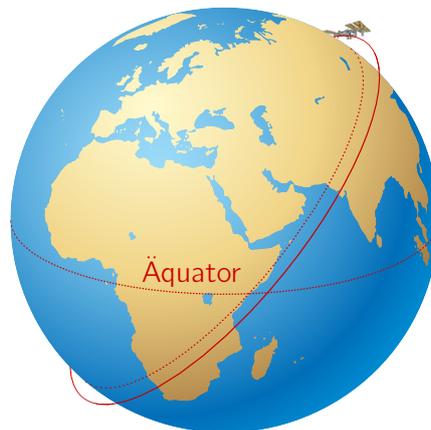

**Abbildung 5:** Orbit der Internationalen Raumstation über der Erde. Die Höhe des Orbits ist im selben Maßstab wie der Erddurchmesser eingezeichnet (Bild: M. Nielbock/HdA/NASA).

Um zu verstehen, wie die Bahnhöhe der ISS und ihre Bahngeschwindigkeit zusammenhängen, werden einige Grundkenntnisse der Mechanik benötigt, die nachfolgend dargestellt werden.

## Himmelsmechanik

Die Himmelsmechanik beschreibt die Bewegung von Himmelskörpern – sowohl natürliche als auch künstliche – durch mathematische und physikalische Gesetze. Sie ist eine Spezialdisziplin der klassischen Mechanik. Die Bewegungen von Himmelskörpern entsprechen meistens periodischen Bahnen von Kreisen oder Ellipsen, manchmal auch Parabeln und Hyperbeln. Die häufig in der Schule genutzten kartesischen Koordinaten sind eher für geradlinige Bewegungen geeignet. Bewegungen entlang einer Kurve können in solch einem System rasch kompliziert werden. Ein Beispiel für geradlinige, gleichförmige Bewegungen wird im nachfolgenden Video gezeigt, in dem die italienische Astronautin Samantha Cristoforetti in der ISS mit einigen Bällen jongliert.

How Mass and Gravity Work in Space - Classroom Demonstration (Englisch, Dauer: 9:19 min)
https://youtu.be/Vl6Mrhgum2c

Kreisbewegungen werden in der kartesischen Darstellung wegen den auftretenden, variierenden Winkeln schnell kompliziert. Stattdessen bieten sich ebene oder sphärische Polarkoordinaten an. Hier verwendet man neben Strecken auch Winkel. So durchläuft ein Satellit während eines kompletten Umlaufs einen Winkel von $2\pi$ bzw. $360°$. Das Beispiel in Abb. 6 erläutert, wie ein Punkt sowohl in kartesischen als auch in Polarkoordinaten beschreiben werden kann. Er hat die kartesischen Koordinaten $x = 4$ und $y = 3$. Derselbe Punkt lässt sich jedoch dadurch Beschreiben, dass man seinen Abstand vom Koordinatenursprung $r$ sowie den Winkel bezüglich einer Richtung − hier die x-Achse − angibt. Hier wird als Winkel $\varphi$ angegeben. Damit ist derselbe Punkt erneut eindeutig bestimmt.

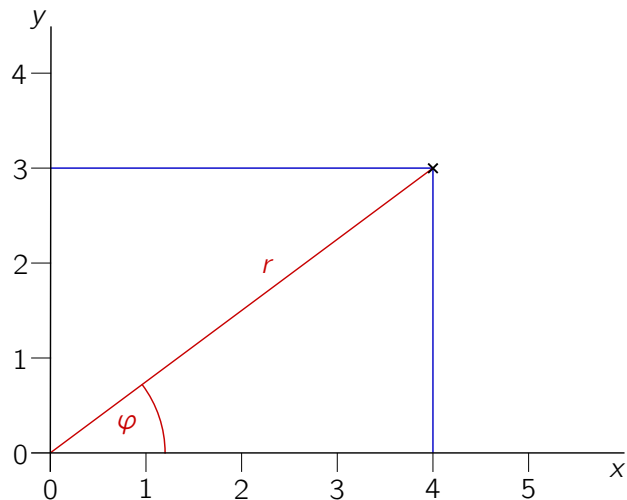

**Abbildung 6:** Illustration der Beziehung zwischen kartesischen Koordinaten und Polarkoordinaten (Grafik: M. Nielbock/HdA).

Da $r$ das Rechteck, das durch $x$ und $y$ erzeugt wird, in zwei rechtwinklige Dreiecke aufteilt, gilt:

$$\frac{y}{r} = \sin\varphi \tag{1}$$

$$\frac{x}{r} = \cos\varphi \tag{2}$$

$$\frac{y}{x} = \tan\varphi \tag{3}$$

$$r^2 = x^2 + y^2 \tag{4}$$

In diesem Beispiel mit $x = 4$ und $y = 3$ finden wir somit:

$$r = \sqrt{4^2 + 3^2} = \sqrt{16 + 9} = \sqrt{25} = 5$$

$$\varphi = \arctan\frac{3}{4} = 0,6435 = 36,87°$$

Rotiert der Punkt in einem Kreis um den Koordinatenursprung, bleibt $r$ konstant, und $\varphi$ definiert die Position auf der Bahn. Mit kartesischen Koordinaten sind beide Koordinaten variabel und über Winkelfunktionen miteinander verknüpft.

Laut erstem newtonschen Axiom behält ein Körper seinen Bewegungszustand ohne äußere Krafteinwirkung bei, d. h. sowohl seine Geschwindigkeit als auch seine Richtung. Bei einer Kreisbewegung handelt es sich um einen Vorgang, bei dem ein Objekt durch eine ständig wirkende Kraft von seiner geradlinigen Bewegung auf eine entsprechende Bahn gezwungen wird.

Im Falle von Satelliten ist das die Gravitation. Sie wirkt senkrecht zur Bewegungs- bzw. Geschwindigkeitsrichtung. Solch eine Kraft wird Zentripetalkraft genannt. Sie übt eine Zentripetalbeschleunigung auf den Satelliten aus. Dabei wird allerdings nicht der Betrag der Geschwindigkeit verändert, sondern lediglich die Richtung. Da diese Kraft permanent wirkt, ergibt sich eine periodische Kreisbewegung.

## Zentripetalbeschleunigung – mit Skalaren

Eine einfache Darstellung für die plausible mathematische Ableitung der Zentripetalbeschleunigung ist über die Grundgleichungen der Mechanik sowie ein rechtwinkliges Dreieck möglich (Abb. 7).

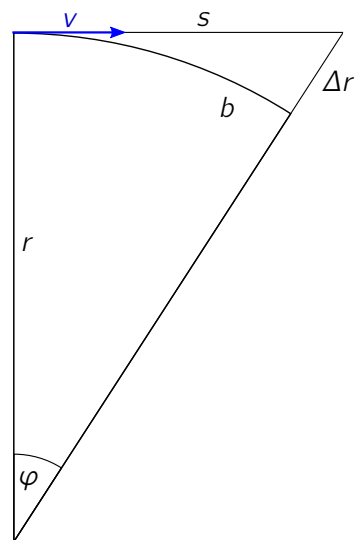

**Abbildung 7:** Einfache Darstellung zur Herleitung der Zentripetalbeschleunigung durch den Vergleich einer geradlinigen Bewegung mit einer Kreisbewegung (Grafik: M. Nielbock/HdA).

Ohne äußere Krafteinwirkung bleibt die Bewegung in Richtung von $v$ geradlinig. Daher legt der Satellit die Strecke $s = v \cdot \Delta t$ zurück. Wirkt jedoch eine Kraft auf den Satelliten ein, wird $v$ zur Bahngeschwindigkeit der resultierenden Kreisbewegung, die senkrecht zur Kraftwirkung verläuft. Die entsprechende Winkelgeschwindigkeit ist $\omega$. Daher legt der Satellit in diesem Fall die Strecke des Kreisbogens $b$ zurück.

$$b = \varphi \cdot r = \omega \cdot \Delta t \cdot r = \omega \cdot r \cdot \Delta t = v \cdot \Delta t \tag{5}$$

Der Wegunterschied $\Delta r$ zwischen $s$ und $b$ wird durch die zusätzliche Beschleunigung $a$ hervorgerufen. Der Betrag der Geschwindigkeit ändert sich dadurch nicht. Somit kann man für $\Delta r$ schreiben:

$$\Delta r = \frac{1}{2} a \cdot \Delta t^2 \tag{6}$$

Damit bekommt man ein rechtwinkliges Dreieck mit den folgenden Seitenlängen:

$$r \tag{7}$$

$$s = v \cdot \Delta t \tag{8}$$

$$r + \Delta r = r + \frac{1}{2} a \cdot \Delta t^2 \tag{9}$$

Hinweis! Dies ist nur eine Näherung, denn tatsächlich sind $s = r \cdot \tan \varphi$ und $b = r \cdot \varphi$ nicht gleich lang. Da am Ende jedoch $\varphi \to 0$ angenommen wird, ist die Differenz zwischen $s$ und $b$ beliebig klein und geht für den Grenzübergang gegen Null. Daraus folgt mit dem Satz von Pythagoras:

$$r^2 + (v \cdot \Delta t)^2 = \left( r + \frac{1}{2} a \cdot \Delta t^2 \right)^2 \tag{10}$$

$$\Leftrightarrow r^2 + v^2 \cdot \Delta t^2 = r^2 + 2 \cdot r \cdot \frac{1}{2} a \cdot \Delta t^2 + \left( \frac{1}{2} a \cdot \Delta t^2 \right)^2 \tag{11}$$

$$\Leftrightarrow v^2 \cdot \Delta t^2 = r \cdot a \cdot \Delta t^2 + \frac{1}{4} a^2 \cdot \Delta t^4 \tag{12}$$

$$\Leftrightarrow v^2 = r \cdot a + \frac{1}{4} a^2 \cdot \Delta t^2 \tag{13}$$

Nun soll aber nicht $v$ für ein beliebiges $\varphi$ sondern am Startpunkt der Bewegung ermittelt werden. Das heißt, $\varphi$ wird sehr klein. Man kann hier auch argumentieren, dass die Wirkdauer der Beschleunigung sehr kurz gewählt werden muss, weil man sonst über eine zu große Zeitdauer mittelt. Somit wird auch $\Delta t$ klein. Für $\Delta t \to 0$ verschwindet der hintere Term. Es bleibt die Zentripetalbeschleunigung:

$$v^2 = r \cdot a \tag{14}$$

$$\Leftrightarrow a = \frac{v^2}{r} = \omega^2 \cdot r \tag{15}$$

### Zentripetalbeschleunigung – mit Vektoren (für Fortgeschrittene)

Will man dies auf etwas höherem Niveau herleiten, benötigt man Vektoren und hier insbesondere die Darstellung durch Einheitsvektoren. Bei der Transformation einer allgemeinen Bewegung von kartesischen in Polarkoordinaten tauchen neue Terme auf, die die Kreisbewegungen von Orbits anschaulich beschreiben. Zentralkräfte wie die durch die Gravitation hervorgerufene Zentripetalkraft werden offensichtlich. Solche Zentralkräfte sind es, die gemäß dem Newtonschen Prinzip *Actio = Reactio* eine sich gleichförmig bewegende Masse von ihrer geraden Bahn ablenken. Die Masse wird ständig in Richtung des Kraftzentrums beschleunigt und auf eine Kreisbahn gezwungen.

Abbildung 8 zeigt die Bewegung eines Massenpunkts $m$, wobei er mit der Bahngeschwindigkeit $\vec{v}_t$ um eine Achse dreht, die über die Winkelgeschwindigkeit $\vec{\omega}$ definiert ist. Um nun die Bewegung von $m$ zu beschreiben, bieten sich anstatt den kartesischen Koordinaten Polarkoordinaten an. Das Koordinatensystem hat in dieser Darstellung seinen Ursprung am Anfang des Ortsvektors $\vec{r}$, der vom Zentrum der Drehbewegung ausgeht. Neben dem Abstand $|\vec{r}|$ des Massenpunkts $m$ vom Koordinatenursprung hat dieses Koordinatensystem die Koordinate $\varphi$. Will man nun in diesen Koordinaten die Geschwindigkeit von $m$ errechnen, muss man den Ortsvektor nach der Zeit ableiten.

$$\vec{v} = \dot{\vec{r}} = \frac{\mathrm{d}}{\mathrm{d}t} \vec{r} = \frac{\mathrm{d}}{\mathrm{d}t} (r \cdot \vec{e_r}) \tag{16}$$

Hier ist $\vec{e_r}$ der Einheitsvektor in $r$-Richtung. Mit der Produktregel folgt:

$$\vec{v} = \dot{r} \cdot \vec{e_r} + r \cdot \dot{\vec{e_r}} \tag{17}$$

$$= v_r \cdot \vec{e_r} + r \cdot \dot{\vec{e_r}} \tag{18}$$

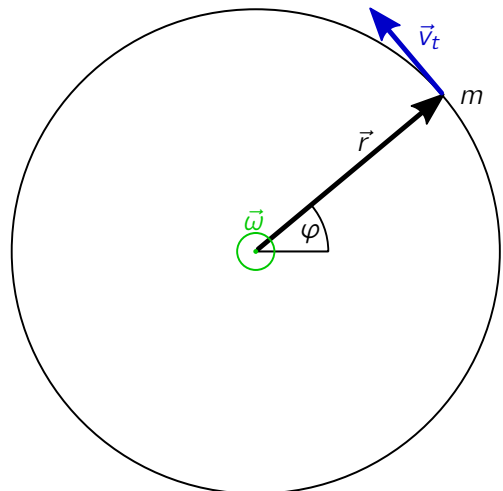

**Abbildung 8:** Vektordarstellung der Rotation eines Massenpunkts $m$ in einem Inertialsystem mit dem Ortsvektor $\vec{r}$, dem Geschwindigkeitsvektor der Rotation $\vec{v}_t$ und dem Vektor der Rotation $\vec{\omega}$ (Grafik: M. Nielbock/HdA).

Somit sieht man, dass die Geschwindigkeit von $m$ in zwei Komponenten zerfällt. Die erste ist die Radialkomponente, die eine Geschwindigkeit in Richtung des Ortsvektors darstellt. Die wird bei reinen Rotationsbewegungen jedoch vernachlässigt. Hinzu kommt eine Komponente mit einer zeitlichen Ableitung von $\vec{e}_r$. Um zu verstehen, was diese zeitliche Ableitung bedeutet, betrachte man Abb. 8. Bei einer Rotation mit $\vec{\omega}$ bewegt sich $m$ in der Rotationsebene auf einem Kreis.

Die einzige Änderung, die $\vec{e}_r$ vollziehen kann, ist seine Richtung. Der Betrag eines Einheitsvektors kann sich nicht ändern. Aber er kann sich drehen. Die Differenz der Richtungen von $\vec{e}_r$ zwischen zwei Zeitpunkten kann man wiederum durch einen Vektor $\vec{\Delta e}_r$ darstellen (Abb. 9) mit:

$$\vec{\Delta e}_r = \vec{e}_r(t + \Delta t) - \vec{e}_r(t) \tag{19}$$

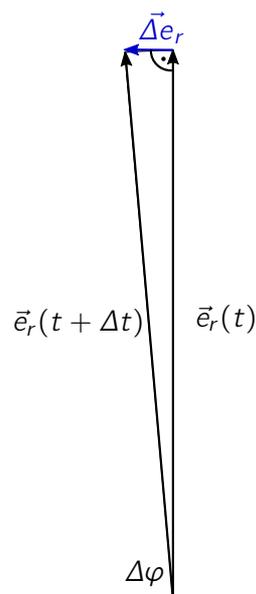

**Abbildung 9:** Drehung des Einheitsvektors $\vec{e}_r$ um $\vec{\Delta e}_r$ innerhalb von $\Delta t$ in Richtung des Einheitsvektors $\vec{e}_\varphi$ (Grafik: M. Nielbock/HdA).

In guter Näherung kann man für kleine Änderungen von $\Delta t$ bzw. $\Delta \varphi$ die Beziehungen zwischen diesen Vektoren wie in Abb. 9 als rechtwinkliges Dreieck darstellen. Da $\vec{\Delta e}_r$ in einem rechten Winkel zu $\vec{e}_r$ steht, verläuft er in Richtung des Einheitsvektors der $\varphi$-Koordinate, also $\vec{e}_\varphi$. Somit kann man ansetzen:

$$\vec{\Delta e}_r = \alpha \cdot \vec{e}_\varphi \tag{20}$$

Den Betrag $\alpha$ von $\vec{\Delta e}_r$ erhält man durch die Beziehungen in einem rechtwinkligen Dreieck.

$$\frac{|\vec{\Delta e}_r|}{|\vec{e}_r|} = \tan(\Delta \varphi) \tag{21}$$

$$\Leftrightarrow \frac{\alpha \cdot |\vec{e}_\varphi|}{|\vec{e}_r|} = \tan(\Delta \varphi) \tag{22}$$

$$\Leftrightarrow \alpha = \tan(\Delta \varphi) \tag{23}$$

Damit lässt sich die Frage beantworten, was der Betrag der zeitlichen Änderung des Einheitsvektors in $r$-Richtung ist.

$$|\dot{\vec{e}}_r| = \left.\frac{\vec{\Delta e}_r}{\Delta t}\right|_{\Delta t \to 0} = \left.\frac{\tan(\Delta \varphi)}{\Delta t}\right|_{\Delta t \to 0} \tag{24}$$

Für kleine $\Delta \varphi$ gilt $\tan(\Delta \varphi) = \Delta \varphi$. Somit folgt:

$$|\dot{\vec{e}}_r| = \left.\frac{\tan(\Delta \varphi)}{\Delta t}\right|_{\Delta t \to 0} = \left.\frac{\Delta \varphi}{\Delta t}\right|_{\Delta t \to 0} = \dot{\varphi} \tag{25}$$

Dadurch erhält man mit Gl. 18:

$$\vec{v} = v_r \cdot \vec{e}_r + r \cdot \dot{\varphi} \cdot \vec{e}_\varphi \tag{26}$$

$$= v_r \cdot \vec{e}_r + r \cdot \omega \cdot \vec{e}_\varphi \tag{27}$$

Das ist das erwartete Ergebnis, denn $r \cdot \omega$ ist nichts anderes als der Betrag der Tangentialgeschwindigkeit $\vec{v}_t$, also $|\vec{v}_t| = v_t$. In anderen Worten: Bei einer Drehbewegung dreht sich der Massenpunkt $m$ mit einer Bahn- bzw. Tangentialgeschwindigkeit $\vec{v}_t$ um den Kreismittelpunkt. Dabei ist der Betrag dieser Geschwindigkeit das Produkt aus der Winkelgeschwindigkeit $\omega$ und dem Kreisradius $r$.

Als nächstes wird nun die Beschleunigung in Polarkoordinaten berechnet.

$$\ddot{\vec{r}} = \dot{\vec{v}} = \frac{\mathrm{d}}{\mathrm{d}t}(\dot{r} \cdot \vec{e}_r) + \frac{\mathrm{d}}{\mathrm{d}t}(r \cdot \dot{\varphi} \cdot \vec{e}_\varphi) \tag{28}$$

$$= \ddot{r} \cdot \vec{e}_r + \dot{r} \cdot \dot{\vec{e}}_r + \dot{r} \cdot \dot{\varphi} \cdot \vec{e}_\varphi + r \cdot (\ddot{\varphi} \cdot \vec{e}_\varphi + \dot{\varphi} \cdot \dot{\vec{e}}_\varphi) \tag{29}$$

$$= \ddot{r} \cdot \vec{e}_r + \dot{r} \cdot \dot{\varphi} \cdot \vec{e}_\varphi + \dot{r} \cdot \dot{\varphi} \cdot \vec{e}_\varphi + r \cdot \ddot{\varphi} \cdot \vec{e}_\varphi + r \cdot \dot{\varphi} \cdot \dot{\vec{e}}_\varphi \tag{30}$$

$$= \ddot{r} \cdot \vec{e}_r + (2 \cdot \dot{r} \cdot \dot{\varphi} + r \cdot \ddot{\varphi}) \cdot \vec{e}_\varphi + r \cdot \dot{\varphi} \cdot \dot{\vec{e}}_\varphi \tag{31}$$

Es ergeben sich also Terme, die sowohl in radialer als auch in tangentialer Richtung wirken. Es bleibt nun noch, $\dot{\vec{e}}_\varphi$ zu ermitteln. Der Ansatz für die Lösung besteht darin, dass aus Abb. 9 ersichtlich ist, dass $\Delta \vec{e}_r$ und damit $\vec{e}_\varphi$ stets senkrecht auf $\vec{e}_r$ steht. $\vec{e}_\varphi$ vollzieht also die Drehung von $\vec{e}_r$ um $\Delta \varphi$ mit. Daher befindet sich auch $\dot{\vec{e}}_\varphi$ bzw. $\vec{\Delta e}_\varphi$ senkrecht auf $\dot{\vec{e}}_r$, also senkrecht auf $\vec{e}_\varphi$ und somit parallel zu $-\vec{e}_r$ (Abb. 10).

---

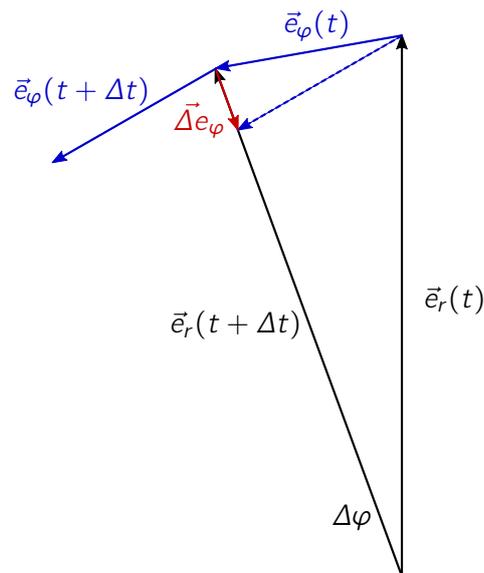

**Abbildung 10:** Drehung des Einheitsvektors $\vec{e}_\varphi$ um $\vec{\Delta e}_\varphi$ in Richtung des Einheitsvektors $-\vec{e}_r$ (Grafik: M. Nielbock/HdA).

In Anlehnung an Gl. 20 kann man schreiben:

$$\vec{\Delta e}_\varphi = \beta \cdot (-\vec{e}_r) \tag{32}$$

Hier gilt wegen der gleichförmigen Drehung:

$$\beta = \alpha = \Delta\varphi \tag{33}$$

Daraus folgt:

$$\dot{\vec{e}}_\varphi = \frac{\vec{\Delta e}_\varphi}{\Delta t} = \frac{-\Delta\varphi \cdot \vec{e}_r}{\Delta t} = -\dot\varphi \cdot \vec{e}_r \tag{34}$$

Eingesetzt in Gl. 31 erhält man:

$$\ddot{\vec{r}} = \ddot{r} \cdot \vec{e}_r + (2 \cdot \dot{r} \cdot \dot\varphi + r \cdot \ddot\varphi) \cdot \vec{e}_\varphi - r \cdot \dot\varphi^2 \cdot \dot{\vec{e}}_r \tag{35}$$

$$= (\ddot{r} - r \cdot \dot\varphi^2) \cdot \vec{e}_r + (2 \cdot \dot{r} \cdot \dot\varphi + r \cdot \ddot\varphi) \cdot \vec{e}_\varphi \tag{36}$$

**Tabelle 2:** Beschleunigungsterme, die bei allgemeinen Bewegungen ausgedrückt in Polarkoordinaten auftreten können.

| Term | mathematischer Ausdruck | Richtung |
|---|---|---|
| Radialbeschleunigung | $\ddot{r}$ | positive $r$-Richtung |
| Zentripetalbeschleunigung | $r \cdot \dot\varphi^2$ | negative $r$-Richtung |
| Coriolisbeschleunigung | $2 \cdot \dot{r} \cdot \dot\varphi$ | positive $\varphi$-Richtung (tangential) |
| Eulerbeschleunigung | $r \cdot \ddot\varphi$ | positive $\varphi$-Richtung (tangential) |

Somit spaltet sich die Gesamtbeschleunigung ausgedrückt in Polarkoordinaten in Terme auf, wie sie in Tab. 2 dargestellt sind. Die gesuchte Zentripetalbeschleunigung weist also zum Zentrum der Kreisbewegung. Es gilt dieselbe Beziehung wie in Gl. 15. Damit wird der Bahnradius festgelegt, der sich mit der Bahngeschwindigkeit eines Objekts einstellt. Da die Zentripetalbeschleunigung von

Satelliten durch die Gravitationsbeschleunigung hervorgerufen wird, lässt sich für jedes $v_t$ bzw. $\omega$ ein dazu assoziierter Bahnradius berechnen. Umgekehrt folgt aus einer vorgegebenen Bahnhöhe eines Satelliten die Bahngeschwindigkeit, die erreicht werden muss, um eine konstante Kreisbahn zu erreichen und einzuhalten.

## Analogie: Kanonenkugel

Eine Satellitenbahn ist tatsächlich ein Spezialfall eines Wurfs parallel zur Erdoberfläche. Wenn man von einem hohen Berg eine Kanone abfeuert, hängt die erreichte Weite von der Geschwindigkeit ab, auf die man die Kugel horizontal – also parallel zur Erdoberfläche – beschleunigt. Am Ende wird sie aber unweigerlich zu Boden fallen. Ist die Geschwindigkeit jedoch groß genug, reicht die Erdbeschleunigung nicht mehr aus, um die Kugel auf eine ausreichend große Geschwindigkeit zum Erdboden hin abzulenken. Die Kugel fällt somit ständig um die Erde herum (Bahnen C und D in Abb. 11).

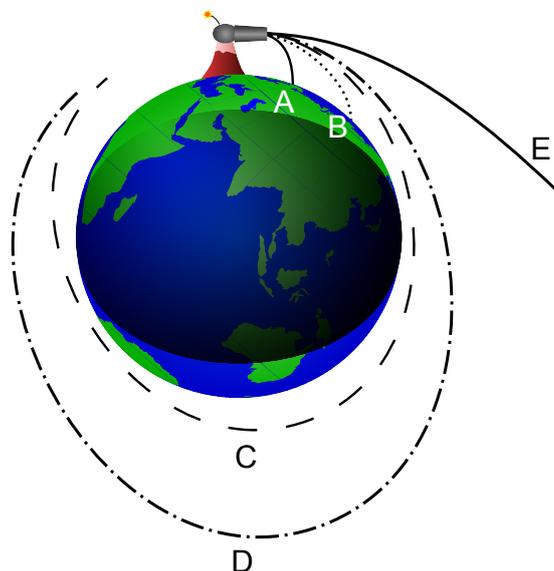

**Abbildung 11:** Ein Gedankenexperiment mit einer horizontal abgeschossenen Kanonenkugel, die bei genügend großer Anfangsgeschwindigkeit um die Erde kreist (Bild: Brian Brondel (https://commons.wikimedia.org/wiki/File:Newton_Cannon.svg), https://creativecommons.org/licenses/by-sa/3.0/legalcode).

## Zentrifugalkraft

Dies waren Betrachtungen, die sich auf einen äußeren Beobachter eines sich bewegenden Objekts beziehen. Begibt man sich in das System des bewegten Objekts, so nimmt man die eigene Bewegung nicht wahr. Ohne äußere Referenzpunkte lässt sich nicht beurteilen, ob man sich bewegt oder nicht. Und selbst mit solchen Punkten lässt sich nicht ohne weiteres sagen, ob sich die Punkte bewegen oder man sich selbst bewegt. Allerdings treten innerhalb eines beschleunigten Bezugssystems Kräfte auf, die scheinbar keine erkennbare Ursache haben. Diese werden offenbar, wenn man das beschleunigte System von Außen betrachtet. Von dort sind diese sogenannten Scheinkräfte ein Resultat der Massenträgheit des beschleunigten Objekts.

Ein mitbewegter Beobachter erfährt eine Beschleunigung, die ihn radial nach außen drängt. Tatsächlich liegt aber eine Beschleunigung hin zum Zentrum der Kreisbewegung vor, die Zentripetalbeschleu-nigung. Vom Betrag sind Zentripetal- und Zentrifugalbeschleunigung identisch. Sie wirken jedoch in entgegengesetzte Richtungen. Auf eine mathematische Herleitung der Zentrifugalbeschleunigung wird hier verzichtet. Nach der Transformation der Koordinaten von einem äußeren, feststehenden Koordinatensystem in des beschleunigte Referenzsystem sind die mathematischen Berechnungen sehr ähnlich wie bei der Ableitung der Zentripetalkraft. Eine ausführliche Herleitung findet man beispielsweise in dem folgenden Video.

Scheinkräfte/Corioliskraft/Zentrifugalkraft/Beschleunigendes Bezugssystem (Dauer: 24:14 min)
https://youtu.be/1kpea6vhy2g

Aus dem beschleunigten, d. h. rotierenden Bezugssystem heraus lässt sich ebenfalls eine Beziehung zur anziehenden Gravitation herstellen. Das Argument lautet demnach, dass die Zentrifugalkraft für eine bestimmte Bahngeschwindigkeit in einem zu berechnenden Abstand vom Gravitationszentrum die wirkende Schwerkraft kompensiert. Es stellt sich ein Kräftegleichgewicht bzw. Schwerelosigkeit ein. Es gilt daher:

$$\vec{F}_{zf} = -\vec{F}_g \tag{37}$$

$$\Rightarrow \left|\vec{F}_{zf}\right| = \left|-\vec{F}_g\right| \tag{38}$$

$$\Leftrightarrow F_{zf} = F_g \tag{39}$$

## Bahnhöhe und Bahngeschwindigkeit von künstlichen Satelliten

Da die Zentrifugalkraft und die Zentripetalkraft vom Betrag gleich sind, ist die Berechnung des Verhältnisses von Bahngeschwindigkeit, Bahnhöhe und gravitierender Zentralmasse äquivalent, obwohl der argumentative Ansatz jeweils ein anderer ist. Man erhält also:

$$F_g = F_z \tag{40}$$

$$\Leftrightarrow \frac{G \cdot M \cdot m}{r^2} = \frac{m \cdot v^2}{r} \tag{41}$$

$$\Leftrightarrow \frac{G \cdot M}{r} = v^2 \tag{42}$$

$$\Rightarrow v = \sqrt{\frac{G \cdot M}{r}} \tag{43}$$

Für diese Beziehung gilt die Identität der Zentripetalkraft mit der Erdgravitation für einen Massepunkt $m$, der sich im freien Fall befindet. Äquivalent heben sich die Gravitationskraft und die Zentrifugalkraft gegenseitig auf. Allerdings wirkt die Gravitationskraft weiter. Das folgende Video thematisiert dies ebenfalls:

Die drei kosmischen Geschwindigkeiten − Astrophysik (Dauer: 7:44 min)
https://youtu.be/-2o8_JvniDM

Gleichung 43 lässt sich nun beliebig für verschiedene Satelliten anwenden, um ihre Bahnparameter auf ihrem Orbit um die Erde zu bestimmen.

## Zusätzliche Einflüsse auf die Bahn der ISS

Die bisherigen Betrachtungen gehen von idealen Voraussetzungen aus. In Wirklichkeit ist die Physik der Satellitenbahnen viel komplexer.

Das Schwerefeld der Erde ist nicht homogen. Das bedeutet, dass die Gravitation entlang einer Bahn gleichbleibender Höhe leicht schwankt. Das wirkt sich auf den realen Orbit der ISS aus. So sind weder die Bahnhöhe noch die Geschwindigkeit exakt konstant.

Weiterhin befinden sich in 400 km Höhe über der Erde noch einige Luftteilchen der Atmosphäre. Durch die Reibung bremsen sie die ISS und andere Satelliten. Daher nimmt die Geschwindigkeit der ISS mit der Zeit langsam ab und die Raumstation sinkt langsam. Deswegen nutzt man die Triebwerke der angedockten Raumschiffe regelmäßig, um die Bahngeschwindigkeit und die Bahnhöhe zu korrigieren.

## Sonderfall geostationäre Bahn

Eine Reihe von Satelliten erfüllen Aufgaben, bei denen es notwendig ist, dass sie von einem gegebenen Ort auf der Erde stets sichtbar sind. So scheinen viele Telekommunikationssatelliten am Himmel still zu stehen. Nur so kann eine starre Ausrichtung beispielsweise auf Fernsehsatelliten gewährleistet werden. Solche Satelliten werden auch für die Kommunikation zwischen der ISS und der Erde benutzt (Campbell 2015).

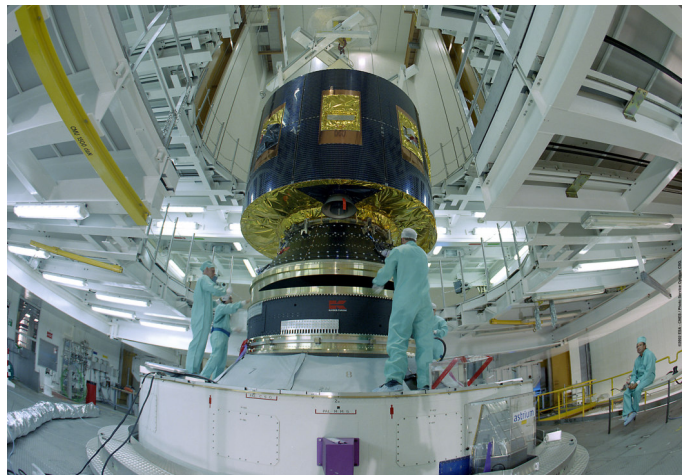

**Abbildung 12:** Der Wettersatellit MSG-1 bei der Integration in die Trägerrakete (Bild: ESA/CNES-Service Optique CSG).

Aber auch Wettersatelliten (Abb. 12), die stets dieselbe Region auf der Erde überwachen, befinden sich relativ zur Erdoberfläche immer an derselben Position (DWD 2018). Die Bahn solcher Satelliten wird geostationärer Orbit genannt (Abb. 13).

Charakteristisch daran ist, dass die Umlaufperiode der Satelliten derjenigen der Rotation der Erde gleicht. Die Rotationsperiode der Erde beträgt 23 Stunden 56 Minuten und 4,099 Sekunden und wird siderischer Tag genannt. Er unterscheidet sich vom bürgerlichen Tag, der sich an der Position der Sonne am Himmel orientiert. Da die Erde während eines Jahres um die Sonne umläuft, ist er etwas länger.

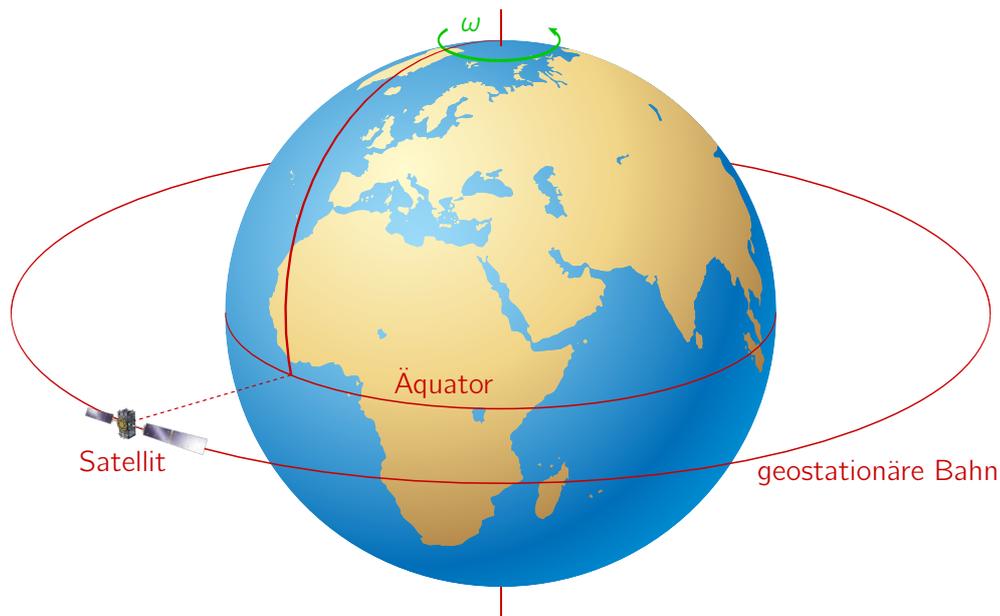

**Abbildung 13:** Position eines geostationären Satelliten über dem Erdäquator (nicht maßstäblich, Bild: M. Nielbock/HdA; ESA/ATG medialab).

Mit Gl. 43 und den Beziehungen

$$\omega = \frac{2 \cdot \pi}{T} \tag{44}$$

$$v = \omega \cdot r = \frac{2 \cdot \pi \cdot r}{T} \tag{45}$$

kann man den Bahnradius $r$ ermitteln, der sich aus der Rotationsperiode $T$ der Erde ergibt. Die Bahnhöhe $h$ folgt aus der Differenz zwischen dem Bahnradius $r$ und dem äquatorialen Erdradius $r_{E0} = 6378$ km. Man erhält:

$$h = r - r_E = \sqrt[3]{\frac{G \cdot M_E \cdot T^2}{4 \cdot \pi^2}} - r_E = 35786 \text{ km} \tag{46}$$

## Aktivität:

### Vorbereitung für Lehrpersonen

Wägen Sie den Schwierigkeitsgrad der Erläuterung der Zentripetalkraft bzw. der Zentrifugalkraft gemäß der Gruppe aus Schülerinnen und Schülern her. Leiten Sie daran angepasst diese Kräfte zusammen mit den Schülerinnen und Schülern ab. Die Motivation kann verbessert werden, indem Sie bereits am Anfang darauf hinweisen, dass Sie Kreisbewegungen wie die der Raumstation im Detail erfassen und mathematisch analysieren wollen. Daraus wird schnell ersichtlich, dass Bewegungen in kartesischen Koordinaten wenig sinnvoll sind, da sie eher an geradlinige Bewegungen angepasst sind. Kreisbewegungen haben dagegen ein Symmetriezentrum, das als Koordinatenursprung dient.

Stellen Sie entsprechende Arbeitsblätter und die in den Aufgaben vorgeschlagenen Videos bereit. Falls Sie sie projizieren möchten, benötigen Sie zumindest einen Computer mit Internetzugang und einen Bildschirm bzw. einen Projektor mit Leinwand.

### Thematische Einführung (Vorschlag)

Für eine atmosphärische und thematische Einführung bieten sich stimmungsvolle Videos wie das folgende an, welches einen Blick von der Internationalen Raumstation auf die Erde wirft.

4K Video from the ISS, April 2016 (Dauer: 1:06 min und 1:37 min)
https://svs.gsfc.nasa.gov/30771

Beim Schauen dieses Videos werden einige Fragen aufgeworfen. Aus welchem Grund fliegt die ISS um die Erde? Warum bleibt sie nicht einfach immer am selben Punkt über der Erde stehen? So wäre sie doch einfacher zu erreichen. Was könnte der Grund für die Geschwindigkeit sein, die die ISS zu einem bestimmten Augenblick besitzt? Wovon hängt die Wahl der Geschwindigkeit ab? Stellen Sie diese Fragen zur Diskussion.

Geben Sie zu bedenken, was mit einem Stein passiert, den man senkrecht in die Luft wirft. Vergleichen Sie diesen Vorgang mit dem waagerechten Wurf, wie er in den Hintergrundinformationen beschrieben wird. Als Erklärungsansatz können auch – mit leichten Abstrichen – folgende Videos dienen.

How Mass and Gravity Work in Space - Classroom Demonstration (Englisch, Dauer: 9:19 min)
https://youtu.be/Vl6Mrhgum2c

Kreisbewegung - Warum fallen Satelliten nicht zurück auf die Erde? (Dauer: 5:33 min)
https://youtu.be/yOZnTmFWwJE

### Aufgaben

Die Aufgaben umfassen folgende Themen:

- Geschwindigkeit der ISS

- Umlaufdauer der ISS

- Fragen zum Verständnis von Satellitenbahnen

- geostationäre Bahn

**Tabelle 3:** Wichtige physikalische Größen und ihre Werte.

| Größe | Formelzeichen | Zahlenwert und Einheit |
|---|---|---|
| Gravitationskonstante | $G$ | $6,67408 \cdot 10^{-11} \; \frac{\mathrm{m}^3}{\mathrm{kg \cdot s^2}}$ |
| Erdmasse | $M_E$ | $5,9722 \cdot 10^{24} \; \mathrm{kg}$ |
| Mittlerer Erdradius | $r_E$ | $6,371 \cdot 10^6 \; \mathrm{m}$ |
| Äquatorialer Erdradius | $r_{E0}$ | $6,378 \cdot 10^6 \; \mathrm{m}$ |
| Bürgerlicher Tag | $T_b$ | $86400 \; \mathrm{s}$ |
| Siderischer Tag | $T_s$ | $86164,099 \; \mathrm{s}$ |

Für die Bearbeitung der Aufgaben sind die Angaben in Tab. 3 hilfreich. Der siderische Tag orientiert sich am scheinbaren Lauf der Sterne (lat. sidus) am Himmel. Die tatsächliche Position der Erde relativ zu den Sternen verändert sich dabei praktisch nicht. Daher entspricht der siderische Tag der Rotationsperiode der Erde. Der bürgerliche Tag orientiert sich am täglichen Sonnenlauf. Im zeitlichen Mittel erscheint die Sonne alle 24 Stunden wieder in derselben Richtung am Himmel. Weil die Erde neben ihrer Rotation zudem während eines Jahres ein Mal um die Sonne läuft, verschiebt sich die scheinbare Position der Sonne gegenüber den Sternen langsam. Die Erde muss daher noch ein wenig weiter rotieren, damit die Sonne wieder in derselben Richtung wie am Vortag erscheint. Daher ist der siderische Tag etwas kürzer als der bürgerliche Tag.

## 1. Die Bahn der ISS

Die Erde zieht alle Objekte durch ihre Gravitation an. Das gilt auch für die ISS. Wirft man einen Stein hoch, fällt er herunter. Ebenso fallen Gegenstände aus großen Höhen zu Boden. Diskutiere mit deinen Mitschülerinnen und Mitschülern, warum das mit der ISS nicht geschieht, obwohl sie doch von der Erde angezogen wird.

## 2. Wie hoch ist die Geschwindigkeit der ISS?

Die ISS umrundet die Erde auf einer Kreisbahn. Auf sie wirken verschiedene Kräfte ein. Dies sind insbesondere die Gravitationskraft $F_g$ und die Zentripetalkraft $F_z$.

$$F_g \quad = \quad \frac{G \cdot M \cdot m}{r^2} \qquad (47)$$

$$F_z \quad = \quad \frac{m \cdot v^2}{r} \qquad (48)$$

Trage die auf die ISS wirkenden Kräfte in Abb. 14 ein.

Betrachte die Kräfteverhältnisse, die die ISS auf einer stabilen Bahn halten. Zeige, dass mit Gl. 40 hieraus eine Bestimmungsgleichung für die Kreisbahngeschwindigkeit folgt.

$$v = \sqrt{\frac{G \cdot M}{r}} \qquad (49)$$

Berechne hieraus die Bahngeschwindigkeit der ISS unter der Annahme, dass sie sich 400 km über der Erdoberfläche befindet. Beachte, dass sich Gl. 49 auf den Bahnradius bezüglich des Erdmittelpunkts bezieht. Bedenke, dass in den Gleichungen die Strecken in Metern angegeben werden müssen. Recherchiere, z. B. auf https://www.dlr.de/next/, wie gut dieser Wert mit den veröffentlichten Daten übereinstimmt.

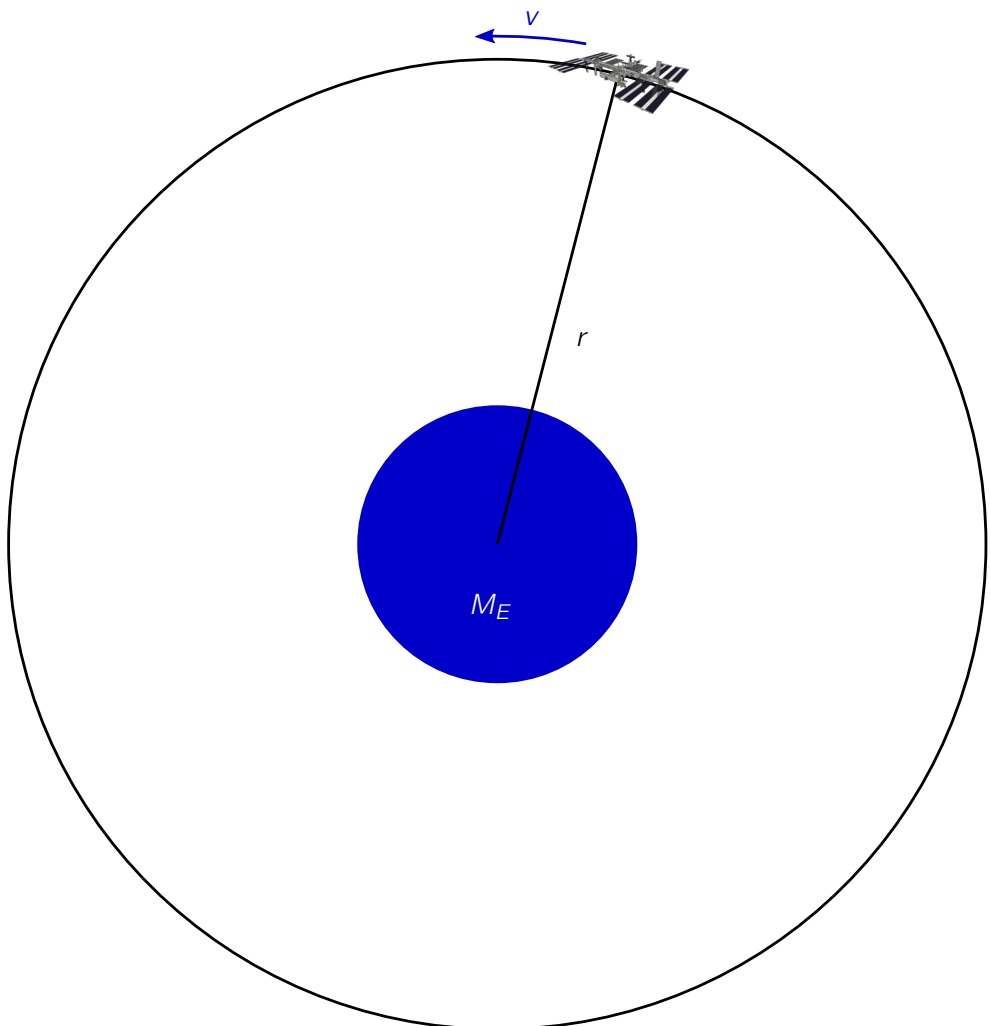

**Abbildung 14:** Modell des Orbits der ISS.

### 3. Wie lange braucht die ISS für eine Erdumkreisung?

Zwischen der Bahngeschwindigkeit $v$, und dem Bahnradius $r$ gilt folgende Relation.

$$v = \frac{2 \cdot \pi \cdot r}{T} \qquad (50)$$

Benutze Gl. 50, um zu berechnen, wie lange die ISS für eine Umrundung der Erde benötigt. Nimm erneut eine Bahnhöhe von 400 km an. Wie viele Umrundungen schafft die ISS am Tag?

### 4. Geostationäre Satelliten (Bonusaufgabe für Fortgeschrittene)

Wir haben gesehen, dass die ISS, wie viele andere Erdsatelliten, die Erde mit hoher Geschwindigkeit mehrmals am Tag umrundet. Eine besondere Gruppe von Satelliten umkreist die Erde über dem Äquator genau synchron mit der Erddrehung. Dieses sind die geostationären Satelliten. Geostationär bedeutet, dass diese Satelliten relativ zu einem Beobachter auf der Erde am Himmel still zu stehen scheinen.

Recherchiere einige geostationäre Satelliten und beschreibe ihre Funktion.

Die Beziehung zwischen Umlaufzeit und Erdrotation stellt sich für einen bestimmten Bahnradius ein. Berechne den Radius, die Höhe der Bahn sowie die Geschwindigkeit von geostationären Satelliten. Beachte, dass die Rotationsdauer der Erde etwas kürzer als ein gewöhnlicher Tag ist und siderischer Tag genannt wird. Verwende dazu die Gleichungen 49 und 50, um zunächst den Bahnradius zu bestimmen.

## 5. Natürliche Satelliten (Bonusaufgabe für Fortgeschrittene)

Auch natürliche Satelliten wie der Erdmond gehorchen denselben Gesetzmäßigkeiten. Berechne mit der Lösung aus Aufgabe 4 die Entfernung des Mondes von der Erde. Der Mond benötigt für einen Umlauf um die Erde die Zeit (d: bürgerliche Tage):

$$T_M = 27,3217 \, \text{d} \tag{51}$$

## Lösungen:

### 1. Die Bahn der ISS

Wie auch schon zuvor kann man dieses Phänomen mit dem freien Fall um die Erde als auch mit dem Gleichgewicht zwischen Gravitationskraft und Zentrifugalkraft erklären. In beiden Erklärungen ist die Geschwindigkeit der ISS die Ursache.

### 2. Wie hoch ist die Geschwindigkeit der ISS?

Die Herleitung entspricht der von Gl. 43. Daraus folgt:

$$v = \sqrt{\frac{G \cdot M}{r}} = \sqrt{\frac{G \cdot M_E}{r_E + h}}$$

$$= \sqrt{\frac{6,67408 \cdot 10^{-11} \frac{m^3}{kg \cdot s^2} \cdot 5,9722 \cdot 10^{24} \, kg}{6371 \, km + 400 \, km}}$$

$$= 7672,5 \, \frac{m}{s} = 7,7 \, \frac{km}{s} = 27621 \, \frac{km}{h}$$

### 3. Wie lange braucht die ISS für eine Erdumkreisung?

Mit Gl. 45 kann man eine Beziehung zwischen der Bahngeschwindigkeit $v$ und der Umlaufperiode $T$ herstellen. Daraus folgt:

$$v = \frac{2 \cdot \pi \cdot r}{T} = \sqrt{\frac{G \cdot M_E}{r}} \tag{52}$$

$$\Leftrightarrow T^2 = \frac{4 \cdot \pi^2 \cdot r^3}{G \cdot M_E} \tag{53}$$

$$\Rightarrow T = 2 \cdot \pi \cdot \sqrt{\frac{r^3}{G \cdot M_E}} \tag{54}$$

Hinweis! Gl. 53 entspricht dem 3. Keplerschen Gesetz.

$$\Rightarrow T = 2 \cdot \pi \cdot \sqrt{\frac{(r_E + h)^3}{G \cdot M_E}}$$

$$= 2 \cdot \pi \cdot \sqrt{\frac{(6371 \, km + 400 \, km)^3}{6,67408 \cdot 10^{-11} \frac{m^3}{kg \cdot s^2} \cdot 5,9722 \cdot 10^{24} \, kg}}$$

$$= 5544,9 \, s = 92,4 \, min$$

### 4. Geostationäre Satelliten

Typische Funktionen von geostationären Satelliten sind die Erdbeobachtung und die Fernerkundung sowie die Analyse des Wetters. Weitere Anwendungsgebiete sind Kommunikation und Fernsehen.

Die Höhe der Bahn wird in Gl. 46 hergeleitet. Die Geschwindigkeit von geostationären Satelliten lässt sich durch die Beziehung aus Gl. 45 ermitteln.

$$
\begin{aligned}
v &= \frac{2 \cdot \pi \cdot r}{T_s} = \frac{2 \cdot \pi \cdot (r_{E0} + h)}{T_s} \\
&= \frac{2 \cdot \pi \cdot (6378\,\text{km} + 35786\,\text{km})}{86164{,}099\,\text{s}} \\
&= 3{,}1\,\frac{\text{km}}{\text{s}}
\end{aligned}
$$

## 6. Natürliche Satelliten

Mit Gl. 53 und durch Einsetzen der Zahlenwerte erhält man:

$$
\begin{aligned}
r &= \sqrt[3]{\frac{G \cdot M_E \cdot T^2}{4 \cdot \pi^2}} \\
&= 383180\,\text{km}
\end{aligned}
$$

# Literatur

## Danksagung

Der Autor bedankt sich bei den Lehrern Matthias Penselin, Florian Seitz und Martin Wetz für ihre wertvollen Hinweise, Kommentare und Änderungsvorschläge, die in die Erstellung dieses Materials eingeflossen sind. Weiterer Dank gilt Herrn Dr. Volker Kratzenberg-Annies für seine gewissenhafte Durchsicht.

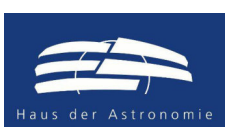
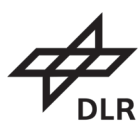
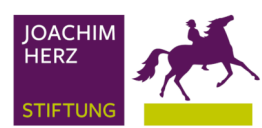